\documentclass[12pt]{article}
\usepackage{bbm}
\newcommand{\be}{\begin{equation}}
\newcommand{\ee}{\end{equation}}
\newcommand{\ba}{\begin{eqnarray}}
\newcommand{\ea}{\end{eqnarray}}
\newcommand{\no}{\nonumber\\}
\begin{document}
\title{\normalsize \hfill UWThPh-2003-9 \\[1cm] \LARGE
A non-standard $CP$ transformation \\
leading to maximal atmospheric neutrino mixing}
\author{Walter Grimus\thanks{E-mail: walter.grimus@univie.ac.at} \\
\setcounter{footnote}{3}
\small Institut f\"ur Theoretische Physik, Universit\"at Wien \\
\small Boltzmanngasse 5, A--1090 Wien, Austria \\*[3.6mm]
Lu\'{\i}s Lavoura\thanks{E-mail: balio@cfif.ist.utl.pt} \\
\small Universidade T\'ecnica de Lisboa \\
\small Centro de F\'\i sica das Interac\c c\~oes Fundamentais \\
\small Instituto Superior T\'ecnico, P--1049-001 Lisboa, Portugal \\*[4.6mm] }

\date{10 October 2003}

\maketitle

\begin{abstract}
We discuss a neutrino mass matrix $\mathcal{M}_\nu$ originally found
by Babu, Ma, and Valle (BMV) and show that this mass 
matrix can be characterized by a simple algebraic relation. From this
relation it follows that atmospheric neutrino mixing is
exactly maximal while at the same time an arbitrary mixing angle
$\theta_{13}$ of the lepton mixing matrix $U$ is allowed and---in the
usual phase convention---$CP$ violation in mixing is maximal; 
moreover, neither the
neutrino mass spectrum nor the solar mixing angle are restricted. We
put forward a
seesaw extension
of the Standard Model,
with three right-handed neutrinos and three Higgs doublets, where 
the family lepton numbers are softly broken by the Majorana mass terms of
the right-handed neutrino singlets and the BMV mass matrix results
from a non-standard $CP$ symmetry.
\end{abstract}

\newpage

\section{Introduction}

The atmospheric neutrino problem,
with mixing angle $\theta_{23}$,
requires $\sin^2 2\theta_{23} > 0.92$ at 90\% CL,
with a best fit value $\sin^2 2\theta_{23} = 1$,
i.e.\ maximal mixing \cite{SK}.
There are many models and textures in the literature
which attempt to explain large---not necessarily
maximal---atmospheric neutrino mixing---for reviews see Ref.~\cite{reviews}.
But,
the closer the experimental lower bound on $\sin^2 2\theta_{23}$ comes to 1,
the more urgent it becomes to find a rationale
for \emph{maximal} atmospheric mixing.
Unfortunately this is not an easy task.
Maximal mixing means $\left| U_{\mu 3} \right| = \left| U_{\tau 3} \right|$,
where $U$ is the lepton mixing matrix,
and this in general requires a $\mu$--$\tau$ interchange symmetry,
which on the other hand must be broken since $m_\mu \neq m_\tau$. 
For a recent discussion of this point see Ref.~\cite{volkas}.

Two models for maximal atmospheric mixing have been suggested by us,
one of them \cite{GL01} based on lepton-number symmetries
softly broken at the seesaw scale,
the other one \cite{GL03} based on a discrete symmetry
spontaneously broken at the same scale.
Both models yield an effective mass matrix for the light left-handed neutrinos
at the seesaw scale
\be
\mathcal{M}_\nu = \left( \begin{array}{ccc} x & y & y \\ y & z & w \\
y & w & z \end{array} \right),
\label{jhgui}
\ee
where $x$, $y$, $z$, and $w$ are in general complex.
The matrix
\be
H = \mathcal{M}_\nu^\ast \mathcal{M}_\nu
\ee
then has an eigenvector $\left( 0,\ 1,\ -1 \right)^T$,
and therefore the models predict $U_{e3} = 0$
besides $\left| U_{\mu 3} \right| = \left| U_{\tau 3} \right|$;
they will have to be discarded if $\left| U_{e3} \right|$
is experimentally found to be non-zero.

A different approach has been suggested by Babu, Ma, and Valle (BMV) \cite{ma}.
Starting from a degenerate neutrino mass matrix at the seesaw scale,
and using the renormalization group
in the context of softly broken supersymmetry
with a general slepton mass matrix,
they have obtained at the weak scale
\be
\mathcal{M}_\nu = \left( \begin{array}{ccc} a & r & r^\ast \\ r & s & b \\
r^\ast & b & s^\ast \end{array} \right),
\label{urtye}
\ee
where $r$ and $s$ are in general complex while $a$ and $b$ remain real.
BMV have found that,
under some approximations,
this $\mathcal{M}_\nu$ yields maximal atmospheric mixing and,
furthermore,
an imaginary $U_{e3}$ (``maximal $CP$ violation'')
in the standard phase convention (to be specified shortly).

It is important to note that the mass matrix in Eq.~(\ref{urtye})
is \emph{not} a generalization of the one in Eq.~(\ref{jhgui}).
If $r$ and $s$ in Eq.~(\ref{urtye}) are real,
then that mass matrix coincides with the one in Eq.~(\ref{jhgui})
with $x$, $y$, $z$, and $w$ real.
On the other hand,
the mass matrix of Eq.~(\ref{urtye}) in general yields
a non-zero $U_{e3}$ as soon as $r$ and $s$ are complex,
while (\ref{jhgui}) always yields $U_{e3} = 0$,
even when $x$, $y$, $z$, and $w$ are complex.

In Ref.~\cite{harrison},
Harrison and Scott (HS) suggested that the lepton mixing matrix may satisfy
\be
\label{absolute}
\left| U_{\mu j} \right| = \left| U_{\tau j} \right|
\quad {\rm for} \quad j = 1,2,3\, ,
\ee
which is still a particular case of maximal atmospheric mixing,
but has the advantage of being more general than the extra condition
$U_{e3} = 0$.
HS showed that,
if $U$ satisfies Eq.~(\ref{absolute}) and the neutrinos are Dirac particles,
then $U$ may be parametrized as
\be
U = \left( \begin{array}{ccc} u_1 & u_2 & u_3 \\ w_1 & w_2 & w_3 \\
w_1^\ast & w_2^\ast & w_3^\ast \end{array} \right),
\label{HSmatrix}
\ee
where the $u_j$ $ (j = 1, 2, 3)$ are real and non-negative,
while the $w_j$ are complex and satisfy the orthogonality conditions
\be
2\, \mathrm{Re} \left( w_j w_k^\ast \right) = \delta_{jk} - u_j u_k\, .
\ee
HS also introduced
the concept of ``$\mu$--$\tau$ reflection,''
which they defined as ``the combined operation of $\mu$--$\tau$
flavor exchange [...] and $CP$ transformation on the leptonic sector''
and which is embodied in the mixing matrix of Eq.~(\ref{HSmatrix}).

It is the purpose of this
paper to,
firstly,
prove that the mass matrix of BMV
\emph{always} yields Eq.~(\ref{absolute}) and,
as a consequence,
\emph{exact} maximal atmospheric neutrino mixing and maximal $CP$ violation.
We shall also show that the BMV mass matrix leads to a mixing matrix
of the form in Eq.~(\ref{HSmatrix}),
even while the neutrinos are Majorana particles.
Secondly,
we shall put forward a
model,
based on softly broken lepton numbers
and on the non-standard $CP$ symmetry
called ``$\mu$--$\tau$ reflection'' by HS,
which obtains the BMV mass matrix
at the seesaw scale---without the need for the renormalization group,
for supersymmetry,
or for an extended fermion spectrum like in the original BMV model.
In this way we conclude that maximal atmospheric mixing
is compatible with a non-zero $U_{e3}$
and can be obtained in
an
extension of the Standard Model.

\section{The mass matrix}

Let $\mathcal{M}_\nu$ be a symmetric complex $3 \times 3$ matrix,
the Majorana mass matrix of the light neutrinos,
defined by
\be
\mathcal{L}_{\rm mass}
= \frac{1}{2}\, \nu_L^T C^{-1} \mathcal{M}_\nu \nu_L + {\rm H.c.}
\ee
($C$ is the Dirac--Pauli charge conjugation matrix),
in the basis where the charged-lepton mass matrix is diagonal.
The lepton mixing matrix $U$ is the diagonalizing matrix of $\mathcal{M}_\nu$,
defined by
\be\label{V}
U^T \mathcal{M}_\nu U = \hat m \equiv \mathrm{diag}
\left( m_1, m_2, m_3 \right),
\ee
where the masses $m_j$ are real and non-negative.

\noindent
\textbf{Lemma:} Suppose $U$ and $U'$ satisfy Eq.~(\ref{V})
and the masses are non-degenerate.
Then there is a diagonal unitary matrix $X$ such that $U' = U X$.
Furthermore,
$X_{jj}$ is an arbitrary phase factor if $m_j = 0$,
while $X_{jj} = \pm 1$ for $m_j \neq 0$.

\noindent \textbf{Proof:} Since both $U$ and $U'$ fulfill Eq.~(\ref{V}),
$\mathcal{M}_\nu = U^* \hat m U^\dagger = {U'}^* \hat m {U'}^\dagger$,
or
\be
\label{W}
W^* \hat m = \hat m W \quad \mathrm{with} \quad W = {U'}^\dagger U\, .
\ee
This equation,
together with the non-degeneracy of the masses,
forces $W$ to be diagonal, i.e.\ $W^* = X$.
It is moreover clear that $W_{jj}$ is real when $m_j$ is non-zero,
Q.E.D.

The matrix $\mathcal{M}_\nu$ of Eq.~(\ref{urtye}) is \emph{characterized} by
\be\label{S}
S \mathcal{M}_\nu S = \mathcal{M}_\nu^* \quad \mathrm{with} \quad
S = \left( \begin{array}{ccc} 1 & 0 & 0 \\ 0 & 0 & 1 \\ 0 & 1 & 0
\end{array} \right).
\ee
Let us write $U = ( c_1, c_2, c_3 )$ with column vectors $c_j$.
Equation~(\ref{V}) means that
\be
\mathcal{M}_\nu c_j = m_j c_j^\ast\, .
\label{sityr}
\ee
Starting from this equation and using Eq.~(\ref{S})
we see that 
\be
\mathcal{M}_\nu \left( S c_j^\ast \right) = m_j \left( S c_j^\ast \right)^\ast.
\ee
We thus have a second diagonalizing matrix $U' = S U^\ast$.
Using the lemma above we find that,
if the masses are non-degenerate,
\be
\label{Vcond}
S U^\ast = U X\, .
\ee
Consequently,
Eq.~(\ref{absolute}) holds.

An alternative proof of Eq.~(\ref{absolute})
starts from the observation that the matrix $H$
corresponding to the $\mathcal{M}_\nu$ of Eq.~(\ref{urtye}) has
\be
H_{\mu \mu} = H_{\tau \tau} \quad {\rm and} \quad
H_{e \mu} = H_{e \tau}^\ast\, .
\ee
As a consequence,
\be
\left( H^n \right)_{\mu \mu} = \left( H^n \right)_{\tau \tau}
\label{ityru}
\ee
for any positive integer $n$.
Using $H = U {\hat m}^2 U^\dagger$,
it follows from Eq.~(\ref{ityru}) for two distinct values of $n$
that either there are degenerate neutrinos
or Eq.~(\ref{absolute}) holds.

A popular representation of $U$ \cite{PDG} is given by
\be
\label{U}
U =
{\rm diag} \left( e^{i \alpha_1},\ e^{i \alpha_2},\ e^{i \alpha_3} \right)
U_{23} U_{13} U_{12}\,
{\rm diag} \left( 1,\ e^{i \beta_1},\ e^{i \beta_2} \right),
\ee
with
\ba
\label{U23}
U_{23} &=& \left( \begin{array}{ccc} 1 & 0 & 0 \\
0 & c_{23} & s_{23} \\ 
0 & - s_{23} & c_{23} \end{array} \right), \\
\label{U13}
U_{13} &=& 
\left( \begin{array}{ccc} c_{13} & 0 & s_{13} e^{-i \delta} \\
0 & 1 & 0 \\ - s_{13} e^{i\delta} & 0 & c_{13} \end{array} \right), \\
\label{U12}
U_{12} &=&
\left( \begin{array}{ccc} c_{12} & s_{12} & 0 \\
- s_{12} & c_{12} & 0 \\ 0 & 0 & 1 \end{array} \right).
\ea
The phases $\alpha_j\ (j  = 1, 2, 3)$ are unphysical (unobservable);
$\delta$ is the Dirac phase and
$\beta_{1,2}$
are the Majorana phases.
Computing the product of matrices one gets
\be\label{U123}
U_{23} U_{13} U_{12} = \left( \begin{array}{ccc}
c_{12} c_{13} & s_{12} c_{13} & s_{13} e^{-i\delta} \\
- s_{12} c_{23} - c_{12} s_{23} s_{13} e^{i\delta} &
c_{12} c_{23} - s_{12} s_{23} s_{13} e^{i\delta} &
s_{23} c_{13} \\
s_{12} s_{23} - c_{12} c_{23} s_{13} e^{i\delta} &
- c_{12} s_{23} - s_{12} c_{23} s_{13} e^{i\delta} &
c_{23} c_{13} \end{array} \right).
\ee
Equation~(\ref{absolute}) applies to this product.
From that equation with $j = 3$ one obtains
\be\label{sqrt2}
c_{23} = s_{23} = \frac{1}{\sqrt{2}}\, .
\ee
Now we inspect Eq.~(\ref{absolute}) with $j = 1,2$.
We know experimentally that $c_{12} s_{12} \neq 0$,
since solar neutrinos oscillate \cite{goswami}.
It follows that
\be\label{sc}
s_{13} \cos{\delta} = 0\, ,
\ee
i.e.\ either $U_{e3} = 0$ or $CP$ violation is maximal.
Conversely,
if we require maximal $CP$ violation
($e^{i \delta} = \pm i$)
and maximal atmospheric neutrino mixing
($\left| U_{\mu 3} \right| = \left| U_{\tau 3} \right|$)
with the parameterization of Eq.~(\ref{U123}),
it is easy to see that Eq.~(\ref{absolute}) follows \cite{harrison}. 

We stress that the mass matrix of Eq.~(\ref{urtye})
restricts \emph{neither} the neutrino mass spectrum
\emph{nor} the solar neutrino mixing angle $\theta_{12}$.
In the general case $\cos{\delta} = 0$ also $\theta_{13}$ \emph{remains free}.
Note that,
with Eq.~(\ref{sqrt2}),
the parameter measured in atmospheric neutrino oscillations is
$\sin^2 2\theta_\mathrm{atm} = 
4\, |U_{\mu 3}|^2 \left( 1 - |U_{\mu 3}|^2 \right) = 1 - s_{13}^4$.

Now we want to discuss the relation between the BMV mass matrix
and the parameterization of the mixing matrix in Eq.~(\ref{HSmatrix}).
We stick to the---experimentally justified---assumption 
that the neutrinos are non-degenerate,
and we employ again Eq.~(\ref{Vcond}).
If $m_j \neq 0$, 
then we know that the $X_{jj}$ are either $+1$ or $-1$.
If $X_{jj} = +1$ then we see from Eq.~(\ref{Vcond}) that
\be
c_j = \left( \begin{array}{c} u_j \\ w_j \\ w_j^\ast \end{array} \right),
\label{thereal}
\ee
with real $u_j$.
If $X_{jj} = -1$ then
\be
c_j = \left( \begin{array}{c} i u_j \\ w_j \\ - w_j^\ast \end{array} \right).
\label{theimaginary}
\ee
It remains to consider the possibility $m_j = 0$.
In that case we have  $S c_j^* = c_j X_{jj}$
with an arbitrary phase factor $X_{jj}$.
Since the massless case allows rephasing of the Majorana neutrino field,
one can absorb a factor $(X_{jj})^{-1/2}$ into that field.
It is easy to see that,
then,
$c_j$ assumes the form in Eq.~(\ref{thereal}).
In the case of Eq.~(\ref{theimaginary}),
we may multiply the physical neutrino field by a factor $i$,
thereby passing from Eq.~(\ref{theimaginary}) to Eq.~(\ref{thereal}),
but also changing the sign in front of $m_j$ in Eq.~(\ref{sityr}).
We may also,
if needed,
multiply the neutrino fields by factors $-1$
so that all three $u_j$ become non-negative.
We thus obtain the following result:
if the mass matrix is of the BMV type
(and since neutrinos are non-degenerate)
then the lepton mixing matrix $U$ is of the form in Eq.~(\ref{HSmatrix}),
but the Majorana phase factors $\eta_j$ may be either 1 or $i$;
or,
in other words,
with a $U$ of the form in Eq.~(\ref{HSmatrix}), 
the BMV mass matrix is diagonalized as
$U^T \mathcal{M}_\nu U = \mbox{diag}
\left( \eta_1^2 m_1,\ \eta_2^2 m_2,\ \eta_3^2 m_3 \right)$.

Let us finally consider the conditions under which $U_{e3}$ will be zero
with a mass matrix of the BMV type.
Suppose $r^2 s^\ast$ in Eq.~(\ref{urtye}) is real.
Then,
\be
\mathcal{M}_\nu = Y \left( \begin{array}{ccc}
a & \left| r \right| & \left| r \right| \\
\left| r \right| & \pm \left | s \right| & b \\
\left| r \right| & b & \pm \left | s \right|
\end{array} \right) Y
\quad \mathrm{with} \quad
Y = \mathrm{diag} \left( 1\, ,\ e^{i \arg r}\, ,\ e^{- i \arg r} \right).
\ee
This means that $\mathcal{M}_\nu$ is essentially identical
to the matrix in Eq.~(\ref{jhgui}),
and we conclude that \emph{if} $r^2 s^\ast$ \emph{is real then} $U_{e3} = 0$.
Conversely,
let us now suppose that the neutrino masses are non-degenerate. 
Then we know,
from Eq.~(\ref{Vcond}),
that $S c_3^\ast = \pm c_3$.
(This relation,
with the plus sign,
also holds for $m_3 = 0$,
as we have argued in the previous paragraph.)
If $U_{e3} = 0$ this means $c_3 =
\left( 0\, ,\ w_3\, ,\ \pm w_3^\ast \right)^T$.
Now,
from Eq.~(\ref{sityr}),
$\mathcal{M}_\nu c_3 = m_3 c_3^\ast$.
This gives
\ba
r w_3 \pm \left( r w_3 \right)^\ast &=& 0\, , \label{A}
\\
s w_3 \pm b w_3^\ast &=& m_3 w_3^\ast\, . \label{B}
\ea
Equation~(\ref{B}) implies that $s w_3^2$ is real.
Equation~(\ref{A}) implies that $r^2 w_3^2$ is real.
As $w_3 \neq 0$,
we conclude that
\emph{provided the neutrinos are non-degenerate,}
$U_{e3} = 0$ \emph{implies a real} $r^2 s^\ast$.
We have thus shown that,
with the mass matrix of BMV,
$U_{e3}$ being zero is equivalent to $r^2 s^\ast$ being real.

\section{A model}
\label{model}

We now want to produce a
model that leads to the mass matrix of Eq.~(\ref{urtye}).
In doing this we find inspiration
in our model of maximal atmospheric neutrino mixing of Ref.~\cite{GL01}.
Thus,
we supplement the Standard electroweak Model with three right-handed neutrinos
and two extra Higgs doublets.
We denote the three lepton families $e$,
$\mu$,
and $\tau$ by the general index $\alpha$;
thus, we have three left-handed lepton doublets
\be
D_\alpha
= \left( \begin{array}{c} \nu_{\alpha L} \\ \alpha_L \end{array} \right),
\quad \alpha = e, \mu, \tau\, ,
\ee
together with three right-handed charged-lepton singlets $\alpha_R$
and three right-handed neutrino singlets $\nu_{\alpha R}$.
In the scalar sector we employ three Higgs doublets
\be
\phi_j = \left( \begin{array}{c} \varphi_j^+ \\ \varphi_j^0
\end{array} \right),
\quad j = 1, 2, 3\, .
\ee
These Higgs doublets acquire vacuum expectation values (VEVs)
$\left\langle 0 \left| \varphi_j^0 \right| 0 \right\rangle = v_j
\left/ \sqrt{2} \right.$,
and $v = \sqrt{\left| v_1 \right|^2 + \left| v_2 \right|^2 +
\left| v_3 \right|^2} \simeq 246\, {\rm GeV}$ represents the Fermi scale.

We introduce the three $U(1)$ lepton-number symmetries $L_\alpha$.
These symmetries are meant to be broken \emph{only softly}
at the high seesaw scale.
We also introduce a ${\mathbbm Z}_2$ symmetry under which
$\mu_R$,
$\tau_R$,
$\phi_2$,
and $\phi_3$ change sign.
This symmetry ${\mathbbm Z}_2$ is broken \emph{only spontaneously}
by the VEVs $v_2$ and $v_3$.
Because of the lepton-number symmetries
and of the ${\mathbbm Z}_2$ symmetry,
the Yukawa Lagrangian of the leptons is
(see also Refs.~\cite{GL01,GL03})
\ba
\mathcal{L}_{\rm Y} &=&
- \frac{\sqrt{2}}{v_1}
\left( \begin{array}{cc} \varphi_1^0, & - \varphi_1^+ \end{array} \right)
\sum_{\alpha = e, \mu, \tau}
f_\alpha \bar \nu_{\alpha R} D_\alpha
- \frac{\sqrt{2} m_e}{v_1^\ast}
\left( \begin{array}{cc} \varphi_1^-, & {\varphi_1^0}^\ast \end{array} \right)
\bar e_R D_e
\no & &
- \sum_{j=2}^3 \sum_{\alpha = \mu, \tau}
\left( \begin{array}{cc} \varphi_j^-, & {\varphi_j^0}^\ast \end{array} \right)
g_{j \alpha} \bar \alpha_R D_\alpha,
\label{mghjy}
\ea
where the three $f_\alpha$ and the four $g_{j \alpha}$
are complex numbers
($m_e$ is real without loss of generality and represents the electron mass).
Notice that,
through the first line of Eq.~(\ref{mghjy}),
the smallness of the neutrino masses may be correlated
with the smallness of the electron mass.
The $\mathbbm{Z}_2$ above
is analogous to the auxiliary $\mathbbm{Z}_2$ of Refs.~\cite{GL01,GL03}. 

The right-handed neutrinos have Majorana mass terms given by
\be
\mathcal{L}_{\rm M} = \frac{1}{2} \left( \nu_R^T C^{-1} M_R^\ast \nu_R
- \bar \nu_R M_R C \bar \nu_R^T \right),
\label{LM}
\ee
where $M_R$ is a $3 \times 3$ symmetric matrix
in flavor space.
Now,
$M_R$ is not diagonal since the terms in Eq.~(\ref{LM})
have dimension three
and we allow the lepton-number symmetries $L_\alpha$
to be broken softly \cite{GL01}.
Indeed,
$M_R$ is the sole source of lepton mixing in this framework.
According to the seesaw formula \cite{seesaw},
when the eigenvalues of $\sqrt{M_R^\ast M_R}$ are all of order $m_R \gg v$
one has
\be
\mathcal{M}_\nu = - M_D^T M_R^{-1} M_D\, ,
\ee
where $M_D = {\rm diag} \left( f_e,\ f_\mu,\ f_\tau \right)$.
It has been shown in Ref.~\cite{oneloop} that this framework,
in which the tree-level Yukawa couplings are diagonal
but $M_R$ is not,
leads at the one-loop level
to a renormalized theory
with flavor-changing neutral Yukawa interactions,
in which flavor-changing processes like $\mu^\pm \to e^\pm \gamma$
or $Z^0 \to e^\pm \mu^\mp$ are suppressed by inverse powers of $m_R$
while processes like $\mu^\pm \to e^\pm e^+ e^-$
are unsuppressed by any inverse powers of $m_R$---they are suppressed
only by small Yukawa couplings---since they may be mediated
by neutral scalar particles.

We now want to enforce maximal atmospheric neutrino mixing
through an $\mathcal{M}_\nu$ like in Eq.~(\ref{urtye}).
We do this by imposing
the following generalized $CP$ transformation \cite{harrison,ecker}: 
\ba
& &
\nu_{L \alpha} \to i S_{\alpha \beta}
\gamma^0 C \bar \nu_{L \beta}^T\, , \quad
\alpha_L \to i S_{\alpha \beta}
\gamma^0 C \bar \beta_L^T\, ,
\no & &
\nu_{R \alpha} \to i S_{\alpha \beta}
\gamma^0 C \bar \nu_{R \beta}^T\, , \quad
\alpha_R \to i S_{\alpha \beta}
\gamma^0 C \bar \beta_R^T\, ,
\label{cp}
\\ & &
\phi_{1,2} \to \phi_{1,2}^\ast\, , \quad
\phi_3 \to - \phi_3^\ast\, .
\nonumber
\ea
This $CP$ symmetry makes $f_e$ real and $f_\mu = f_\tau^\ast$,
while $g_{2 \mu} = g_{2 \tau}^\ast$ and $g_{3 \mu} = - g_{3 \tau}^\ast$.
Without loss of generality we assume
that $v_1$ is real and positive.
Then we find
\be
M_D^\ast = S M_D S \quad \mathrm{and} \quad M_R^\ast = S M_R S\, ,
\ee
where the second relation
follows from the $CP$ invariance of $\mathcal{L}_\mathrm{M}$.
Therefore $\mathcal{M}_\nu$,
too,
fulfills Eq.~(\ref{S}),
just as we wanted. 

Let us now consider the masses of the $\mu$ and $\tau$ leptons.
Those masses are given by
\be
m_\mu = \frac{1}{\sqrt{2}}
\left| g_{2 \mu} v_2^\ast + g_{3 \mu} v_3^\ast \right|,
\quad
m_\tau
= \frac{1}{\sqrt{2}}
\left| g_{2 \tau} v_2^\ast + g_{3 \tau} v_3^\ast \right|
= \frac{1}{\sqrt{2}}
\left| g_{2 \mu}^\ast v_2^\ast - g_{3 \mu}^\ast v_3^\ast \right|.
\label{mmumtau}
\ee
With the notation $v_{2,3} = \left| v_{2,3} \right| e^{i\vartheta_{2,3}}$,
we obtain 
\be
\label{mutaumasses}
m_\mu = 
\frac{1}{\sqrt{2}} \left| g_{2 \mu} \left| v_2 \right| + 
g_{3 \mu} \left| v_3 \right| 
e^{i \left( \vartheta_2 - \vartheta_3 \right)} \right|,
\quad
m_\tau = 
\frac{1}{\sqrt{2}} \left| g_{2 \mu} \left| v_2 \right| -
g_{3 \mu} \left| v_3 \right| e^{i \left( \vartheta_3 - \vartheta_2 \right)}
\right|.
\ee
Therefore,
$m_\tau \neq m_\mu$ requires
\be
e^{i \left( \vartheta_2 - \vartheta_3 \right)} \neq
- e^{i \left( \vartheta_3 - \vartheta_2 \right)}\, .
\ee

Now let us check the case of $CP$ conservation.
There we have \cite{branco}
\be
\label{CPcons}
v_2^\ast = v_2\, , \quad v_3^\ast = - v_3\, ,
\ee
and therefore
\be
CP\ \mathrm{conservation}\; \Rightarrow \;
e^{i \left( \vartheta_2 - \vartheta_3 \right)} = \pm i\, .
\label{ctuyo}
\ee
Thus,
if $CP$ is conserved we have $m_\mu = m_\tau$;
$CP$ violation is necessary for $m_\mu \neq m_\tau$ 
(for an earlier model of this type see Ref.~\cite{neufeld}).

We have thus shown that,
provided $CP$ is spontaneously broken,
we are able to obtain $m_\mu \neq m_\tau$
while $\mathcal{M}_\nu$ satisfies Eq.~(\ref{S}).
We thus have a model with maximal atmospheric mixing
but a free $\left| U_{e3} \right|$.

\section{Obtaining $m_\mu \ll m_\tau$}

Let us again consider Eq.~(\ref{mmumtau}).
Since $m_\mu$ and $m_\tau$ are both given by essentially
the same VEVs and Yukawa couplings,
it seems natural to expect that they will be of the same order of magnitude,
even if spontaneous $CP$ breaking ensures that they are different.
However, 
in reality one has $m_\mu \ll m_\tau$.
This strong inequality requires in our model
the almost complete cancellation
of two different products of a VEV and a Yukawa coupling.

In general,
if the muon mass is generated
by the Yukawa coupling $\phi^\dagger \mu_R D_\mu$,
where $\phi$ is some Higgs doublet,
then there is a natural explanation for the smallness of $m_\mu$:
one just introduces the symmetry
$\mu_R \to - \mu_R,\ \phi \to - \phi$.
This symmetry in general restricts the Higgs potential
in such a way that it allows a vacuum with
$\left\langle 0 \left| \varphi^0 \right| 0 \right\rangle = 0$.
The muon mass then turns out to be zero.
If one now lets the symmetry $\phi \to - \phi$ be softly broken
by some terms of dimension two in the Higgs potential
($\phi^\dagger \phi^\prime$,
where $\phi^\prime$ is some other Higgs doublet),
then we obtain a technically natural explanation for the smallness of
$\left\langle 0 \left| \varphi^0 \right| 0 \right\rangle$ and thus of $m_\mu$.
Of course,
this only works if $\phi$ does not have any Yukawa couplings
besides the one to the $\mu$,
which is not the case in the Standard Model
since that model only contains one Higgs doublet.

This idea may be implemented in the context of our model
in the previous section.
Let us introduce the extra symmetry \cite{GL03}
\be\label{K}
K: \quad \mu_R \to - \mu_R\, ,\ \phi_2 \leftrightarrow \phi_3
\ee
into that model,
then one obtains $g_{2\mu} = - g_{3\mu}$ and
\be
m_\mu = \frac{\left| g_{2\mu} \right|}{\sqrt{2}}
\left| v_2 - v_3 \right|,
\quad
m_\tau
= \frac{\left| g_{2\mu} \right|}{\sqrt{2}} \left| v_2 + v_3 \right|.
\ee
On the other hand,
the symmetry
$K$ will also restrict the scalar potential,
and this in such a way that there will in general
be a range of parameters of the potential for which the vacuum
leaves that symmetry unbroken,
i.e.\ $v_2 = v_3$.
This immediately leads to $m_\mu = 0$.
(Notice that $v_2 = v_3$ constitutes a maximal spontaneous breaking
of our non-standard $CP$ symmetry,
cf.\ Eq.~(\ref{ctuyo}).)
In order to obtain a non-zero but small $m_\mu$
it is now enough to introduce into the potential
terms which break
$K$ softly.
In our model,
considering its other symmetries $\mathbbm{Z}_2$ and $CP$,
there are two such terms:
\be
\mu_{\mathrm{s}1}
\left( \phi_2^\dagger \phi_2 - \phi_3^\dagger \phi_3 \right)
+
i \mu_{\mathrm{s}2}
\left( \phi_2^\dagger \phi_3 - \phi_3^\dagger \phi_2 \right).
\ee
The constants $\mu_{s1}$ and $\mu_{s2}$ are real. 
A detailed analysis of the potential
is outside the scope of this paper \cite{forthcoming},
but it is intuitive to expect that,
provided $\mu_{\mathrm{s}1}$ and $\mu_{\mathrm{s}2}$ are small,
$\left| v_2 - v_3 \right|$ will also be kept small
and thus $m_\mu \ll m_\tau$ will prevail.
The crucial point is that this inequality
is now protected by a softly-broken symmetry,
and it is therefore technically natural.

\section{Leptogenesis}

Seesaw models offer an attractive possibility
for explaining the observed baryon asymmetry of the universe by leptogenesis
\cite{fukugita,leptogenesis}.
In order to analyse leptogenesis one works in the weak basis
where the Majorana mass matrix $M_R$ of the right-handed neutrinos
is diagonal,
with matrix elements $M_1$,
$M_2$,
and $M_3$.
The Dirac mass matrix $M_D$ is not diagonal in that weak basis.
Consider the Hermitian matrix $R = M_D M_D^\dagger$ in that weak basis.
Then,
the $CP$-violating asymmetry relevant for leptogenesis is
\be
\epsilon \approx - \frac{3 M_1}{16 \pi v^2 R_{11}}\,
\sum_{i=2}^3\, \frac{\mathrm{Im} \left[ \left( R_{1i} \right)^2 \right]}
{M_i}\, ,
\ee
where for simplicity we have assumed $M_1 \ll M_2, M_3$.

In our model,
in the weak basis where $M_D$ is diagonal,
which is the one that we have used before,
one has $M_D = \hat M_D
= \mathrm{diag} ( f_e,\ f_\mu,\ f_\mu^\ast )$
with $f_e$ real,
while $M_R^\ast = S M_R S$.
Now,
we know from Section~2 that a matrix $M_R$
satisfying $M_R^\ast = S M_R S$
is diagonalized by a unitary matrix $V$ of the Harrison--Scott type,
i.e.
\be
V^T M_R V = \hat M_R = \mathrm{diag} \left( M_1\, ,\ M_2\, ,\ M_3 \right)
\quad \mathrm{with} \quad
V = \left( \begin{array}{ccc} p_1 & p_2 & p_3 \\ q_1 & q_2 & q_3 \\
q_1^\ast & q_2^\ast & q_3^\ast \end{array} \right),
\ee
the $p_j$ being real.
We now move to the weak basis where $M_R$ is diagonal
($M_R = \hat M_R$) and $M_D$ is not.
In that weak basis $M_D = V^T \hat M_D$.
Thus,
\be
R = V^T \hat M_D \hat M_D^\dagger V^\ast
= \left( \begin{array}{ccc} p_1 & q_1 & q_1^\ast \\ p_2 & q_2 & q_2^\ast \\
p_3 & q_3 & q_3^\ast \end{array} \right)
\mathrm{diag} \left( f_e^2\, ,\ \left| f_\mu \right|^2\, ,\
\left| f_\mu \right|^2 \right)
\left( \begin{array}{ccc} p_1 & p_2 & p_3 \\ q_1^\ast & q_2^\ast & q_3^\ast \\
q_1 & q_2 & q_3 \end{array} \right).
\ee
It is clear that the matrix $R$ is not only Hermitian but,
as a matter of fact,
real.
Therefore,
\emph{in our model leptogenesis is not possible}.
The root of this fact can be traced directly to the existence 
of a $CP$ symmetry in our model,
which is spontaneously broken only at the weak scale,
i.e.\ much below the scale at which a net 
lepton number is supposed to be generated.
Since $CP$ is unbroken at that super-high scale,
$\epsilon$ is necessarily zero and leptogenesis cannot proceed.

\section{Conclusions}

In this paper we have discussed the mass matrix of Eq.~(\ref{urtye}),
originally found by Babu,
Ma,
and Valle \cite{ma} in the context of a model based on the group $A_4$
and on softly broken supersymmetry with
additional heavy charged-lepton singlets,
an enlarged scalar sector,
and the seesaw mechanism.
In that model,
the relations $a = b$ and $r = s = 0$ hold at the seesaw scale
and the full mass matrix of Eq.~(\ref{urtye})
arises at the weak scale after the renormalization-group evolution
of $\mathcal{M}_\nu$.
(Note, however, that subsequently a much simpler non-supersymmetric
$A_4$ model was proposed by E.\ Ma \cite{ma},
where the radiative corrections to $a = b$ and $r = s = 0$
are generated by an $A_4$ triplet of charged scalars.)

Firstly,
we have shown that the mass matrix (\ref{urtye}) can be
characterized by the algebraic relation in Eq.~(\ref{S}).
If we consider all the parameters of the matrix (\ref{urtye}) as independent,
it follows readily from this characterization
that atmospheric neutrino mixing is maximal
and that either $\theta_{13} = 0$
(and then the $CP$ phase in lepton mixing is physically meaningless),
or $\theta_{13}$ is arbitrary and---in the phase
convention of Eq.~(\ref{U123})---the $CP$ phase is given by $\pi/2$---see
Eq.~(\ref{U123});
this is the more general case.
Moreover,
the neutrino mass matrix of Eq.~(\ref{urtye}) 
fixes neither the neutrino masses nor the solar 
mixing angle.\footnote{As hinted at above,
in the $A_4$ models of Ref.~\cite{ma} the parameters of $\mathcal{M}_\nu$
are not completely independent and,
therefore,
statements about the neutrino mass spectrum and $\theta_{13}$ can be made.}

Secondly,
we have
derived this mass matrix in the context of a
model based on the lepton sector of the Standard Model with
three right-handed neutrinos, the seesaw mechanism, and
three---instead of one---Higgs doublets.
We have constructed our model in two steps:
\begin{enumerate}
\item Inspired by Refs.~\cite{GL01,GL03},
we have imposed the three $U(1)_{L_\alpha}$ symmetries associated with the
family lepton numbers, which are softly broken
by the $\mathcal{L}_\mathrm{M}$ of Eq.~(\ref{LM}).
\item As suggested by the relation (\ref{S}) and by Ref.~\cite{harrison},
we have imposed the non-standard $CP$ symmetry of Eq.~(\ref{cp})
in order to get the neutrino mass matrix of Eq.~(\ref{urtye}).
\end{enumerate}
In this way,
we have obtained a
renormalizable model where lepton mixing
arises solely from the Majorana mass matrix $M_R$ of 
the heavy neutrino singlets and
where $m_\mu \neq m_\tau$ is a consequence of the spontaneous breaking
of the non-standard $CP$ symmetry.

Our model contains a $\mathbbm{Z}_2$ symmetry which is spontaneously
broken at the Fermi scale (see the discussion of
the Yukawa Lagrangian (\ref{mghjy})),
and this may lead to a cosmological problem
through the formation of domain walls at that scale.
An additional symmetry (see Eq.~(\ref{K})) solves the problem of
$m_\mu \ll m_\tau$ in a technically natural way.

We stress that in our model the neutrino masses are completely free,
contrary to what happens in the $A_4$ models
which predict neutrinos to be approximately degenerate.
We also stress that in our model the $\mathcal{M}_\nu$ of Eq.~(\ref{urtye})
holds at the seesaw scale and its form will be slightly changed
by the renormalization-group (RG) evolution down to the Fermi scale,
while in the $A_4$ models Eq.~(\ref{urtye}) holds precisely \emph{after}
the RG evolution. Our
realization of the mass matrix (\ref{urtye})
is an interesting illustration of the fact
that exact maximal atmospheric neutrino mixing
and a non-zero mixing angle $\theta_{13}$ can coexist,
enforced by a symmetry.
This was not the case in the models of Refs.~\cite{GL01,GL03},
where the mass matrix (\ref{jhgui}) was obtained.

\vspace*{10mm}

\noindent \textbf{Acknowledgement} The work of L.L.\ was supported
by the Portuguese \textit{Funda\c c\~ao para a Ci\^encia e a Tecnologia}
under the contract CFIF-Plurianual.

\end{document}